\documentclass[english,aps,preprint]{revtex4}
\usepackage[T1]{fontenc}
\usepackage[latin9]{inputenc}
\usepackage{amsmath}
\usepackage{amssymb}
\usepackage[xdvi]{graphicx}

\makeatletter
\@ifundefined{textcolor}{}
{%
 \definecolor{BLACK}{gray}{0}
 \definecolor{WHITE}{gray}{1}
 \definecolor{RED}{rgb}{1,0,0}
 \definecolor{GREEN}{rgb}{0,1,0}
 \definecolor{BLUE}{rgb}{0,0,1}
 \definecolor{CYAN}{cmyk}{1,0,0,0}
 \definecolor{MAGENTA}{cmyk}{0,1,0,0}
 \definecolor{YELLOW}{cmyk}{0,0,1,0}
 }

\usepackage{amsfonts}\usepackage{babel}\usepackage{dcolumn}\usepackage{bm}\usepackage{amsfonts} \@ifundefined{textcolor}{}{
 \definecolor{BLACK}{gray}{0}
 \definecolor{WHITE}{gray}{1}
 \definecolor{RED}{rgb}{1,0,0}
 \definecolor{GREEN}{rgb}{0,1,0}
 \definecolor{BLUE}{rgb}{0,0,1}
 \definecolor{CYAN}{cmyk}{1,0,0,0}
 \definecolor{MAGENTA}{cmyk}{0,1,0,0}
 \definecolor{YELLOW}{cmyk}{0,0,1,0}
 }

\makeatother

\usepackage{babel}
\begin{document}

\title{Spatially dependent decoherence and anomalous diffussion of quantum
walks}

\author{A. Pérez}

\affiliation{Departament de Física Teòrica and IFIC, Universitat de València-CSIC
\\
 Dr. Moliner 50, 46100-Burjassot, Spain}

\author{A. Romanelli}

\affiliation{Instituto de Física, Facultad de Ingeniería, Universidad de la República,
\\
 C.C. 30, C.P. 11300, Montevideo, Uruguay}
\begin{abstract}
We analyze the long time behavior of a discrete time quantum walk
subject to decoherence with a strong spatial dependence, acting on
one half of the lattice. We show that, except for limiting cases on
the decoherence parameter, the quantum walk at late times behaves
sub-ballistically, meaning that the characteristic features of the
quantum walk are not completely spoiled. Contrarily to expectations,
the asymptotic behavior is non Markovian, and depends on the amount
of decoherence. This feature can be clearly shown on the long time
value of the Generalized Chiral Distribution (GCD).
\end{abstract}
\maketitle

\section{introduction}

Quantum Walks (QW) constitute the quantum analogy to classical Random
Walks. The latter are an important piece in the design of classical
algorithms and are used, for example, to efficiently explore the parameter
space of some model. Here we will consider the special case of a QW
on a line \cite{Aharonov93,Meyer1996,Ambainis2001,ambainis-2003-1,Kempe03,Kendon6,kendon-2006-364}.
As in the classical case, QWs have been proposed as an element to
design quantum algorithms \cite{PhysRevA.67.052307,Ambainis2007,ChildsSTOC200359-68,PhysRevA.70.022314,PhysRevA.78.012310}.
The importance of the QW has increased with the discovery that it
can be used for universal quantum computation \cite{PhysRevLett.102.180501,PhysRevA.81.042330},
and several experimental setups have been proposed or realized to
implement it \cite{Knight2003147,PhysRevA.65.032310,PhysRevA.66.052319,PhysRevA.67.042316,PhysRevA.72.062317,PhysRevA.78.042334,PhysRevA.82.033602,PhysRevLett.104.050502,Xue2009,Zhao2007}.

An important point to be discussed, as in all implementations of quantum
algorithms, is the possible effect of decoherence due to the interaction
of the experimental setup with the environment. This interaction will
change the problem of an isolated quantum device to the one corresponding
to an \textit{open quantum system} \cite{Breuer2007,Weiss2008}. In
most cases, the consequence will be that the quantum algorithm will
loose its quantum advantages and, therefore, its outerperformance
(as compared to classical algorithms) will be ruined. The effect of
decoherence on the QW has been investigated in a number of papers
\cite{PhysRevA.67.032304,Lopez2003,PhysRevA.67.042315,Kendon6,Annabestani2009,Annabestani2010,PhysRevLett.104.153602,Romanelli2010a,Romanelli2010}.
As a general conclusion, it seems clear that decoherence in the coin
space of the walker spoils the performance of the QW more effectively.
It must be noticed, however, that a small amount of decoherence both
on the coin and position of the one dimensional lattice QW might produce
some benefits \cite{PhysRevA.67.042315}, as it produces more uniform
distributions. On the other hand, purely spatial decoherence (i.e.,
decoherence introduced by some kind of defects on the sites of the
lattice) may have a distinctive behavior. For example, the authors
in \cite{Annabestani2010} study the effects of tunneling on the spatial
sites. They find that the characteristic quadratic dependency of the
variance on time is not ruined, even for maximal noise. Also, one
obtains a smooth probability distribution, except for very strong
decoherence.

From the above considerations, it is clear that the effects of noise
on the QW are worth to study in order to design practical quantum
algorithms which use the properties of the QW. Most papers on this
subject have considered the case when the noise appears uniformly
distributed on the lattice. This assumption allows the effects of
decoherence to be treated within a translationally-invariant formalism,
so that some analytical results can be obtained regarding the characteristic
properties of the QW distribution \cite{Annabestani2010,ahlbrecht:042201}.
From the experimental point of view, it is true that one can force
the setup so as to mimic a spatially uniform noise acting on the coin
and reproduce the expected Anderson localization \cite{PhysRevLett.106.180403}.
But the question that arises is what happens if, in a given experiment,
decoherence appears in an uncontrollable, spatially dependent, way.
More precisely, decoherence might appear only on a given region of
the lattice, or it may act stronger in some parts of the system than
in others. Are results developed under the hypothesis of translational
invariance still valid under these circumstances, at least approximately?
Another possibility is that new phenomena, which were not present
for translationally invariant systems, may appear when this restriction
does not apply. We think that examining such a possibility can be
useful for the design of new experiments, and for the general understanding
of decoherence in the QW.

In this paper, motivated by the above discussion, we study a simple
model of non-translationally invariant noise in the QW, in which decoherence
acts on the coin degree of freedom with some probability $p$, but
only when the walker moves on one half of the lattice. As we will
show, even such a simple model will give rise to interesting phenomena,
which were not shown (at least to our knowledge) in previous models.
At first sight, given the interference properties of the problem under
study, one might expect that the characteristic properties of the
QW, such as the quadratic growth of the variance with time, should
be completely destroyed. We show that, in fact, this is not the case,
and we analyze the long time behavior in connection with the given
initial conditions. We observe that, for large time steps, the variance
grows as a power law. Surprisingly, the walk remains subballistic
even for strong values of $p$.

A characteristic property of the QW is the evolution of the Global
Chirality Distribution (GCD) \cite{Romanelli2010a,Romanelli2010,PhysRevA.85.012319}.
This distribution gives information about the chirality of the walk,
independently of the position. It has been found, for example, that
the GCD possesses in some cases an asymptotic limit, which can be
related to the initial conditions of the coin. This result showed
an unexpected behavior of the QW's dynamics, that is more characteristic
of Markov processes. It shows that, watching only the degrees of freedom
associated with the chirality, it would be very hard to appreciate
the unitary character of the quantum evolution. In more generic words,
the simple observation of variables that belong to only one of the
simplest sub-spaces can camouflage the unitary character of the evolution
\cite{Romanelli2010,Castagnino2010}. Therefore, it is clear that
the study of the GCD is necessary in order to understand the equilibrium
between degrees of freedom of the QW. We will study the evolution
of the chirality, as given by the GCD, and specially its long-time
behavior for a QW subject to decoherence for the model described above. 

The paper is organized as follows. In Sect. II we briefly review the
discrete-time QW on a line. In Sect. III we introduce our model for
decoherence, and derive the recursion formulae obeyed by the left
and right components of the GCD. We discuss the asymptotic expressions
for these magnitudes. Numerical calculations that illustrate the behavior
of these magnitudes are shown on Sect. IV. Sect. V summarizes our
main results.

\section{Discrete time QW walk on a line}

The QW may be defined using either its discrete-time or continuous
time version. In this paper, we concentrate on the discrete time version.
The discrete time QW on the line corresponds to the evolution of a
one-dimensional quantum system in a direction which depends on an
additional degree of freedom, the chirality, with two possible states:
{}``left\textquotedblright{}\ $|L\rangle$\ or {}``right\textquotedblright{}\ $|R\rangle$.
The global Hilbert space of the system is the tensor product $H_{s}\otimes H_{c}$
where $H_{s}$ is the Hilbert space associated to the motion on the
line, and $H_{c}$ is the chirality Hilbert space. Let us call $T_{-}$
($T_{+}$) the operators in $H_{s}$ that move the walker one site
to the left (right) of a unidimensional lattice, and $|L\rangle\langle L|$
, $|R\rangle\langle R|$ the chirality projector operators in $H_{c}$.
We consider the unitary transformations 
\begin{equation}
U(\gamma)=T_{-}\otimes|L\rangle\langle L|\text{ }K_{c}(\gamma)+T_{+}\otimes|R\rangle\langle R|\text{ }K_{c}(\gamma),\label{Ugen}
\end{equation}
where $K_{c}(\gamma)=\sigma_{z}\cos\gamma+\sigma_{x}\sin\gamma$,
and $\sigma_{z}$ , $\sigma_{x}$ are Pauli matrices acting in $H_{c}$.
For $\gamma=\pi/4$ the Hadamard coin is obtained. When decoherence
is not present, the unitary operator $U(\gamma)$ evolves the state
in one time step as 
\begin{equation}
|\Psi(t+1)\rangle=U(\gamma)|\Psi(t)\rangle,\label{evolution}
\end{equation}
and the state at time $t$ can be expressed as the spinor 
\begin{equation}
|\Psi(t)\rangle=\sum\limits _{x=-\infty}^{\infty}\left[\begin{array}{c}
a_{x}(t)\\
b_{x}(t)
\end{array}\right]|x\rangle,\label{spinor}
\end{equation}
where the upper (lower) component is associated to the left (right)
chirality.

\section{Decoherence in the QW}

We assume a simple model in which decoherence in the QW appears as
a consequence of the \textit{additional} action of $\sigma_{z}$ on
the coin space with a given probability. Moreover, this operator is
assumed to act only on the semipositive line, with a characteristic
probability $p$. Therefore, the simple dynamics described in the
previous section has to be modified: Instead of Eq. (\ref{evolution}),
one has to deal with a density operator $\rho$ describing the state
of the QW. The dynamics can be described by the action of given Kraus
operators. In our case, we have two operators $E_{1}$ and $E_{2}$,
defined by 
\begin{equation}
E_{1}=\sqrt{1-p\theta(x)}\, U(\gamma),\label{E1}
\end{equation}
 
\begin{equation}
E_{2}=\sqrt{p}\,\theta(x)\sigma_{z}U(\gamma),\label{E2}
\end{equation}
 where $\theta(x)$ is the Heaviside step function. In this way, the
operator $E_{2}$ only acts on a position eigenstate $|x\rangle$
when $x\geq0$, and $E_{1}$ reproduces the ordinary QW, Eq. (\ref{evolution})
whenever $x<0$. In other words, one has the usual QW on the left
side of the lattice, whereas on the right side an additional dephasing
operation $\sigma_{z}$ appears with some probability $p$, where
$p\in[0,1]$ is a real number. One can readily check that the necessary
condition for trace-preserving Kraus operators 
\begin{equation}
E_{1}E_{1}^{\dagger}+E_{2}E_{2}^{\dagger}=1,\label{kraus1}
\end{equation}
 is fulfilled. The time evolution for the density matrix of the quantum
walk is then given by the map 
\begin{equation}
\rho(t+1)=E_{1}\rho(t)E_{1}^{\dagger}+E_{2}\rho(t)E_{2}^{\dagger}.\label{tmap}
\end{equation}
 Later on, we will also consider the map produced on the GCD, which
can be obtained from $\rho(t)$ by tracing out the spatial degrees
of freedom. In this way, we define the reduced density operator for
the chirality evolution 
\begin{equation}
\rho_{c}(t)\equiv Tr_{s}\{\rho(t)\}=\sum_{x}\langle x|\rho(t)|x\rangle.\label{rhoc}
\end{equation}
 From this reduced operator, one can calculate the diagonal components
of the GCD, defined simply as the corresponding elements in the chiral
$\{|L\rangle,|R\rangle\}$ basis

\begin{equation}
\Pi_{L}(t)\equiv\langle L\mid\rho_{c}(t)\mid L\rangle,
\end{equation}
\begin{equation}
\Pi_{R}(t)\equiv\langle R\mid\rho_{c}(t)\mid R\rangle,
\end{equation}

whereas the interference term is given by

\begin{equation}
Q(t)=\frac{1}{2}\left(\langle L\mid\rho_{c}(t)\mid R\rangle+\langle R\mid\rho_{c}(t)\mid L\rangle\right).
\end{equation}

In order to obtain a recursive formula for the GCD, let us expand
the density matrix $\rho(t)$ in the basis $\{|x\rangle\otimes|i\rangle,x\in\mathbb{Z},i=L,R\}$
of the whole Hilbert space, such as 
\begin{equation}
\rho(t)=\sum_{x,y\in\mathbb{Z}}\sum_{i,j=L,R}R_{x,y;i,j}(t)\,\,|x\rangle\langle y|\otimes|i\rangle\langle j|
\end{equation}
 After substitution in Eq. (\ref{rhoc}), using Eqs. (\ref{E1},\ref{E2},\ref{tmap})
we obtain, with the help of some algebra, 
\begin{eqnarray}
Tr_{s}\{E_{1}\rho(t)E_{1}^{\dagger}\} & = & \sum_{x}\{[1-p\theta(x)][M_{R}R_{x,x}(t)M_{R}^{\dagger}+M_{L}R_{x,x}(t)M_{L}^{\dagger}]\nonumber \\
 & + & \sum_{x}\{\sqrt{1-p\theta(x+1)}\sqrt{1-p\theta(x-1)}[M_{R}R_{x-1,x+1}(t)M_{L}^{\dagger}\nonumber \\
 & + & M_{L}R_{x+1,x-1}(t)M_{R}^{\dagger}]\}\label{trE1}
\end{eqnarray}
 where $M_{L}=(|L\rangle\langle L|)K_{c}$ and $M_{L}=(|R\rangle\langle R|)K_{c}$.
On the other hand, 
\begin{eqnarray}
Tr_{s}\{E_{2}\rho(t)E_{2}^{\dagger}\} & = & \sigma_{z}\sum_{x}\{p\theta(x)[M_{R}R_{x,x}(t)M_{R}^{\dagger}+M_{L}R_{x,x}(t)M_{L}^{\dagger}]\nonumber \\
 & + & \sum_{x}\{p\theta(x+1)\theta(x-1)[M_{R}R_{x-1,x+1}(t)M_{L}^{\dagger}\nonumber \\
 & + & M_{L}R_{x+1,x-1}(t)M_{R}^{\dagger}]\}\sigma_{z}.\label{trE2}
\end{eqnarray}
In the above equations, we have introduced 
\begin{equation}
R_{x,y}(t)\equiv\sum_{i,j=L,R}R_{x,y;i,j}(t)\,\,|i\rangle\langle j|,\label{Rxy}
\end{equation}
 which is an operator defined on the coin space. The corresponding
magnitudes at time $t+1$ can be derived from Eqs. (\ref{trE1},\ref{trE2},\ref{Rxy}).
After a lengthy, but straightforward calculation, we arrive to the
following equations relating the diagonal elements of the GCD at time
$t+1$ to those at time $t$:
\begin{equation}
\Pi_{L}(t+1)=\cos^{2}\gamma\,\Pi_{L}(t)+\sin^{2}\gamma\,\Pi_{R}(t)+\sin2\gamma\, Q(t)\label{PILmap1}
\end{equation}
\begin{equation}
\Pi_{R}(t+1)=\sin^{2}\gamma\,\Pi_{L}(t)+\cos^{2}\gamma\,\Pi_{R}(t)-\sin2\gamma\, Q(t)\label{PIRmap2}
\end{equation}

The latter results agree with similar expressions obtained in Refs.
\cite{Romanelli2010a,Romanelli2010,PhysRevA.85.012319} for the GCD
dynamics, but there a handmade technique \cite{Romanelli2003325,Romanelli2004395}
was used to separate the Markovian evolution from the interference
term. It can also be obtained \cite{Romanelli2010} from Eqs. (\ref{PILmap1},\ref{PIRmap2})
that, for $p=0$,  $Q(t)$, $\Pi_{L}(t)$ and $\Pi_{R}(t)$ have long-time
limits whose values are determined by the initial conditions. In the
next section, it is shown that, when $p\neq0$, Eqs. (\ref{PILmap1},\ref{PIRmap2})
also have stationary solutions. 
\begin{figure}[h]

\begin{centering}
\includegraphics[width=8cm]{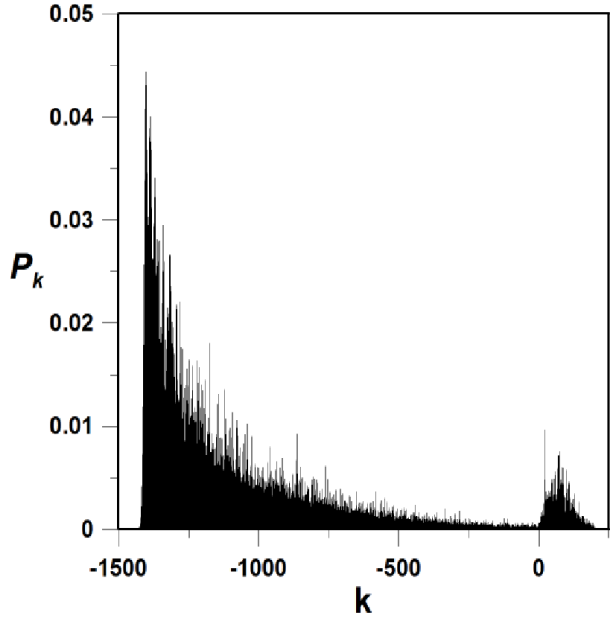} 
\par\end{centering}

\caption{ The position distribution $P_{k}$ as a function of the dimensionless
position at $t=2000$ with $p=0.2$.}

\label{f1} 
\end{figure}

\begin{figure}[h]

\begin{centering}
\includegraphics[width=8cm]{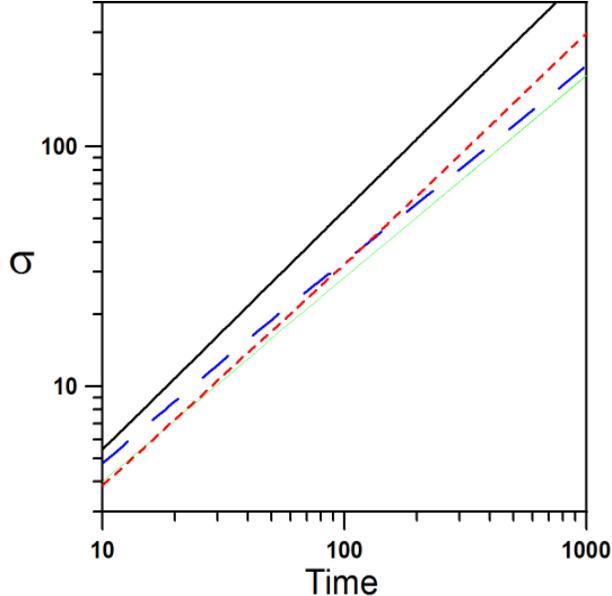} 
\par\end{centering}

\caption{(Color online): The standard deviation, $\sigma$ obtained from all
points on the lattice (positive and negative), as a function of the
time step in log-log scales, for different values of $p$, corresponding
to $p=0$ (black, thick solid line), $p=0.9$ (red, short-dashed line),
$p=0.5$ (green, thin line) and $p=0.1$ (blue, long dashed line).}

\label{f2} 
\end{figure}

\begin{figure}[h]
\begin{centering}
\includegraphics[width=8cm]{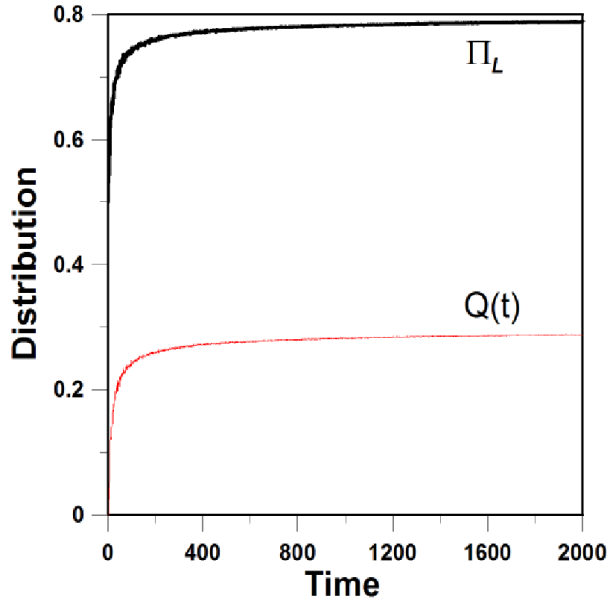} 
\par\end{centering}

\caption{The left component of the GCD and the interference term as functions
of the time step, for $p=0.2$}

\label{f3} 
\end{figure}

According to this, it is expected that the asymptotic GCD will satisfy
the following equations 
\begin{eqnarray}
\Pi_{L}(\infty)=\cos^{2}\gamma\,\Pi_{L}(\infty)+\sin^{2}\gamma\,\Pi_{R}(\infty)+\sin2\gamma\, Q(\infty),\label{PILmap1inf}
\end{eqnarray}
 
\begin{eqnarray}
\Pi_{R}(\infty)=\sin^{2}\gamma\,\Pi_{L}(\infty)+\cos^{2}\gamma\,\Pi_{R}(\infty)-\sin2\gamma\, Q(\infty),\label{PIRmap2inf}
\end{eqnarray}
 where we have defined $\Pi_{L}(\infty)\equiv\Pi_{L}(t\rightarrow\infty)$,
$\Pi_{R}(\infty)\equiv\Pi_{R}(t\rightarrow\infty)$ and $Q(\infty)\equiv Q(t\rightarrow\infty)$.
The asymptotic behavior of the GCD has no explicit dependence on the
parameter $p$, but the asymptotic values of $\Pi_{L}(\infty)$, $\Pi_{R}(\infty)$
and $Q(\infty)$ do have an implicit dependence. Using Eqs. (\ref{PILmap1inf},\ref{PIRmap2inf}),
the stationary solution for the GCD is, then 
\begin{equation}
{\left[\begin{array}{c}
\Pi_{L}(\infty)\\
\Pi_{R}(\infty)
\end{array}\right]}=\frac{1}{2}\left[\begin{array}{c}
1+2Q(\infty)/\tan\gamma\\
1-2Q(\infty)/\tan\gamma
\end{array}\right].\label{estacio}
\end{equation}
 These solutions are the same as those obtained in Ref. \cite{Romanelli2010}
for a more complicated decoherence. Therefore, the dynamical evolution
of the GCD is non Markovian, in the sense that asymptotic magnitudes
depend on the initial conditions (as well as on the probability $p$).
As we show in the next Section, the interference term $Q(\infty)$
is not negligible: The behavior of this composite QW is at first sight
unexpected, since usually decoherence destroys the unitary correlation,
providing a route towards a classical-like behavior described by a
Markov process.

\section{Numerical calculations}

The global evolution of the system depends on the application of the
two operators Eqs. (\ref{E1}, \ref{E2}) in the Hilbert space. Each
operator acts on both the chirality and position spaces, and has a
corresponding map. For our numerical calculations, instead of working
with the density operator, a statistical description can be obtained
by combining many runs of the form Eq. (\ref{evolution}) with the
appropriate statistical weights. Such a description, of course, is
equivalent to working with the density matrix, as long as a sufficiently
large ensemble is considered. For our purposes, we have found that
this procedure is more efficient than a whole calculation involving
a large number of time steps and, consequently, of matrix positions,
if one was to deal with the density matrix. We have checked that the
number of runs in the ensemble is large enough so as to achieve convergence. 

To implement the algorithm we proceed as follows. At each time step
$t$, the usual QW map is applied to each position at the left of
the origin, while at the right half-line the usual QW map, or the
map obtained by the additional action of the Pauli matrix $\sigma_{z}$,
are applied with probabilities $1-p$ and $p$, respectively, $p$
being the only parameter in the model. We take as the initial conditions
a walker starting at the central position $|0\rangle$ with chirality
$\frac{1}{\sqrt{2}}\left(1,i\right)^{T}$, $\gamma=\pi/4$, and we
consider an ensemble of $100$ dynamical trajectories with $2000$
time steps. Finally, magnitudes are averaged over the whole ensemble.

Fig.~\ref{f1} shows the position distribution of the QW given (for
a single run) by $P_{k}(t)\equiv|a_{k}(t)|^{2}+|b_{k}(t)|^{2}$, as
a function of the position $k$ at $t=2000$. The plot, as any other
quantity we will show, reflects the final averaging over the ensemble.
This figure shows, for $p=0.2$, a very different behavior to the
left and to the right of the origin. In particular, the behavior of
the system on the left half-line is close to one of the QW without
decoherence, while on the right half-line the behavior is typical
of a classical walker with a Gaussian distribution. It is worth noticing
that most of the probability goes to the left, as shown in this figure.
In other words, the region on the right, where decoherence is acting,
tends to reflect the walker towards the left, where it can freely
propagate. This effect is important in order to understand other results
that we show below.

As it is well known, one of the most striking properties of the one-dimensional
QW is its ability to spread over the line linearly in time, as characterized
by the standard deviation $\sigma(t)\sim t$, while its classical
analog spreads out as the square root of time $\sigma(t)\sim t^{1/2}$.
In our case, the standard deviation of the system is presented in
Fig.~\ref{f2} for different values of $p$. We see that the standard
deviation grows subballistically, as a power law $\sigma(t)\sim t^{c}$,
where $c$ takes a constant value. It is interesting to note that
this behavior remains for the whole range of values of $p$. For values
of $p$ close to zero, the QW spreads almost ballistically, as expected.
For intermediate values (between $0$ and $1$), most of the probability
goes to the left, and this prevents decoherence from effectively reducing
the QW to a diffusive behavior. Actually, this is only a simplified
vision, as the actual process is more complicated: Indeed, the QW
on the left is actually evolving under the form of waves propagating
\textbf{both} to the left and to the right. These right-moving waves
will penetrate on the right region and eventually be reflected to
the left. It is this complicated interplay between right and left
motions that leads to the subballistic behavior explained above. For
values of $p$ close to unity, the evolution on the right side is
dictated by the operator $\sigma_{z}U(\gamma)$, which is obviously
unitary. Therefore, the QW on both sides will recover the main properties
of the standard QW. 

In Fig.~\ref{f3} we present the left (L) component of the GCD and
the interference term $Q(t)$: Here it is seen that these quantities
have definite limits. We have checked numerically that Eq.(\ref{estacio})
is satisfied independently of both the initial condition and the values
of $p$, for $p>0$.

\section{conclusions}

We have analyzed a model for spatially-dependent decoherence on the
quantum walk, which we implement via the introduction of appropriate
Kraus operators. In our simplified model, the decoherence acts only
on one side of the lattice. We introduced a particular model, defined
as an additional operation acting on the coin, although we have explored
other choices of the Kraus operators, as the ones defined in \cite{Annabestani2009,Annabestani2010}
with similar results. In spite of the noise introduced by these operators,
some characteristic properties of the QW, such as the long term linear
behavior of the standard deviation, still survive. We have calculated
the GCD, and showed that it has a given limit as a function of time.
We conclude that the dynamical evolution of the GCD is not Markovian,
and has an asymptotic value that depends on the initial conditions
and on the probability of decoherence. This is an unexpected result,
since decoherence tends to destroy the correlations arising from the
unitary evolution, thus paving the way to a classical-like Markovian
process. It is also in agreement with other results discussed here,
showing that, in spite of an apparently extreme decoherence, acting
on an semi-infinite lattice, some quantum features are preserved.
\begin{acknowledgments}
This work was supported by the Spanish Grants FPA2011-23897 and 'Generalitat
Valenciana' grant PROMETEO/2009/128. 
\end{acknowledgments}
\bibliographystyle{apsrev}
\bibliography{/home/perez/investig/qcomputing/biblio/books,qwalks}

\end{document}